\documentclass{article}

\usepackage{microtype}
\usepackage{graphicx}
\usepackage{booktabs} 

\usepackage{hyperref}

\usepackage[accepted]{SysML21/mlsys2021}

\usepackage[font=small]{caption}
\usepackage{subcaption}
\usepackage{booktabs}
\usepackage{amsmath}
\usepackage{amsfonts}
\usepackage{listings}
\usepackage{color}
\usepackage{xspace}
\usepackage{comment}
\usepackage{graphicx} 
\usepackage{wrapfig}
\usepackage{multirow}
\usepackage{bm}
\usepackage{microtype}
\usepackage{enumitem}
\usepackage{flushend}
\usepackage{array}
\usepackage{pifont}

\newif\iflong
\longtrue

\setlist{leftmargin=5.5mm}

\definecolor{listinggreen}{rgb}{0,0.6,0}
\definecolor{listinggray}{rgb}{0.5,0.5,0.5}
\definecolor{listingmauve}{rgb}{0.58,0,0.82}
\definecolor{listingkeywordcolor}{rgb}{1.0,0.4,0.0}
\definecolor{listinglightgray}{rgb}{0.8863,0.8863,0.8863}

\newcommand\COMMENT[1]{}
\newcommand\framework{\textsc{Tiramisu}\xspace}

\newcommand{\HIDE}[1]{}

\newcommand{\ModelMAPE}{16}

\newcommand{\ModelPearson}{0.90}
\newcommand{\ModelSpearman}{0.95}
\newcommand{\DataSetSize}{1.8 million\xspace}

\mlsystitlerunning{A Deep Learning Based Cost Model for Automatic Code Optimization}

\begin{document}

\twocolumn[
\mlsystitle{A Deep Learning Based Cost Model for Automatic Code Optimization}





\begin{mlsysauthorlist}
\mlsysauthor{Riyadh Baghdadi}{mit,nyu}
\mlsysauthor{Massinissa Merouani}{esi}
\mlsysauthor{Mohamed-Hicham Leghettas}{esi}
\mlsysauthor{Kamel Abdous}{esi}
\mlsysauthor{Taha Arbaoui}{utt}
\mlsysauthor{Karima Benatchba}{esi}
\mlsysauthor{Saman Amarasinghe}{mit}
\end{mlsysauthorlist}

\mlsysaffiliation{mit}{Massachusetts Institute of Technology}
\mlsysaffiliation{nyu}{New York University Abu Dhabi}
\mlsysaffiliation{esi}{Ecole Nationale Superieure d'Informatique}
\mlsysaffiliation{utt}{University of Technology of Troyes}

\mlsyscorrespondingauthor{Riyadh Baghdadi}{baghdadi@nyu.edu}


\vskip 0.15in

\begin{abstract}
    Enabling compilers to automatically optimize code has been a longstanding goal for the compiler community. Efficiently solving this problem requires using precise cost models. These models predict whether applying a sequence of code transformations reduces the execution time of the program. Building an analytical cost model to do so is hard in modern x86 architectures due to the complexity of the microarchitecture. In this paper, we present a novel deep learning based cost model for automatic code optimization. This model was integrated in a search method and implemented in the \framework{} compiler to select the best code transformations. The input of the proposed model is a set of simple features representing the unoptimized code and a sequence of code transformations. The model predicts the speedup expected when the code transformations are applied. Unlike previous models, the proposed one works on full programs and does not rely on any heavy feature engineering. 
The proposed model has only \ModelMAPE{}\% of mean absolute percentage error in predicting speedups on full programs.
The proposed model enables \framework to automatically find code transformations that match or are better than state-of-the-art compilers without requiring the same level of heavy feature engineering required by those compilers.

\end{abstract}
]



\printAffiliationsAndNotice{}  

\vspace{-0.75cm}
\section{Introduction}

Writing high-performance software is essential in many areas from machine learning to science and engineering. In nuclear physics, for example, researchers need to perform large scale simulations to study the properties of matter. A highly optimized implementation of these simulations can be orders of magnitude faster compared to an unoptimized implementation.
In deep learning, an optimized implementation of a state-of-the-art neural network such as XLNet~\cite{DBLP:journals/corr/abs-1906-08237} is $1.8\times$ faster than the equivalent PyTorch implementation.
Writing such a highly optimized code requires ninja programmers and is time-consuming while the results are error-prone, less understandable, and non-portable. One of the longstanding goals in the compiler community is to develop compilers that can automatically optimize high-level code. These compilers automatically apply code transformations to make the code run faster; thus, avoiding the need for manual low-level program tuning. They provide greater productivity, portability, and high performance, and will be directly accessible by domain scientists. 


Automatically generating efficient code for high-performance systems is a tedious task.
In order for the compiler to generate efficient code, two problems have to be solved. First, a large set of code transformations and a mechanism to apply them to programs need to be provided. Examples of such transformations include loop fission, fusion, parallelization, and vectorization. Second, the right sequence of code transformations from this large set has to be chosen. The selected code transformations must preserve the program semantics and provide the highest performance for the input program.
While state-of-the-art-compilers have shown success in solving the first problem (i.e., the ability to provide a large set of transformations and correctly apply a selected sequence of transformations~\cite{wolf1991loop,bondhugula_practical_2008,trifunovic_graphite_2010,tobias_hexagonal_cgo13,lefebvre_automatic_1998,Qui00}), they still do not successfully solve the second problem (i.e., selecting the sequence of transformations that will provide the best performance).

The problem of selecting the right sequence of code transformations can be modeled as a search problem that can be solved in three steps.
In the first step, the compiler uses a search technique to explore the space of possible code transformations. The result of this step is a set of candidates where each one is a sequence of code transformations. In the second step, the compiler checks the validity of each candidate (i.e., checks that applying the transformations does not change the program semantics).
In the third step, the compiler evaluates the valid candidates and chooses the one that minimizes the execution time. This evaluation can be done by running each candidate on the target hardware to obtain the exact speedup. However, this is not a feasible solution in practice as running a program takes a considerable amount of time.
Moreover, the target hardware may not be available at compile time. Another way to evaluate a candidate is by using a cost model to predict the speedup.

Designing cost models manually is known to be a hard task~\cite{5260526,bachir2013minimal}. This is mainly due to the diversity of hardware architectures and their complexity (out-of-order execution, complex  memory hierarchies, data prefetching, etc.). Complex interactions between code transformations make the problem more complicated.
Recently, cost models, such as Ithemal~\cite{ithemal} and Halide~\cite{Adams:2019:LOH:3306346.3322967}, have demonstrated how to overcome some of this complexity by using deep learning. While these state-of-the-art cost models are more accurate, they are limited in two ways: Ithemal~\cite{ithemal} only predicts throughput for basic blocks of assembly code (instead of full programs). It also assumes that data is always in cache. The cost model in Halide~\cite{Adams:2019:LOH:3306346.3322967} requires heavy feature engineering (it uses 54 complex program features). Designing such features is tedious, error-prone, and time-consuming.

In this paper, we propose a novel DNN-based cost model that avoids the problems of previous work. Our model operates on full programs expressed in a high-level language (not just basic blocks). It takes into consideration not only memory accesses to the cache but also to the main memory. Moreover, it does not require heavy feature engineering.
The proposed cost model takes the original unoptimized code and a sequence of code transformations and predicts the speedup that these transformations would yield when applied. The model is designed for CPUs and is integrated in the \framework{} compiler~\cite{tiramisu}, a compiler for the \framework domain-specific language (DSL). Because this model is a regression model, it allows the compiler to select the best transformation candidates by ranking the candidates selected by a search technique.

\paragraph{Contributions} In summary, the contributions of this paper are:
\begin{itemize}
    \item A novel deep-learning-based cost model for code optimization. This cost model is a \emph{regression} cost model, operates on \emph{full programs}, and \emph{does not rely on extracting complex features}.
    \item An implementation of the proposed model and an integration into a search approach to enable the \framework compiler to automatically search for the best code transformations.
    \item We evaluate the proposed model and show that it has a low error rate reaching \ModelMAPE{}\% mean absolute percentage error. We show also that it enables \framework to automatically find code transformations that match or outperform state-of-the-art compilers.
\end{itemize}

\vspace{-0.35cm}
\section{\framework Embedded DSL}

\framework{}~\cite{tiramisu} is a domain-specific language (DSL) embedded in C++. It provides a C++ API that allows users to write a high level, architecture-independent algorithm, and a set of API calls to select which code transformations should be applied. The first part of a \framework{} program specifies the algorithm without specifying how it should be optimized.
The second part specifies which code transformations to apply and how the results of computations should be stored.
\framework uses a mathematical model known as the polyhedral model internally \cite{feautrier_array_1988,pencil_paper,bondhugula_practical_2008,pencil,tiramisu} to represent code, code transformations, and to reason about the correctness of code transformations.
The following code shows an example of a convolution algorithm written in \framework.

\vspace{-0.25cm}
{\scriptsize
\begin{lstlisting}[language=C,escapechar=@]
// Declare the iterators.
var n(0, batch), fout(0, out_features), fin(0, in_features), y(0, H-2), x(0, W-2), k0(0, 3), k1(0, 3);@\label{fig:example:tiramisu:iterators}@
// Algorithm.
conv(n, fout, y, x) += weights(fout, fin, y, x) * input(n, fin, y + k0, x + k1);@\label{fig:example:tiramisu:computation1}@
\end{lstlisting}
}

The iterators in line~\ref{fig:example:tiramisu:iterators} define the loop bounds around the \texttt{conv} computation. The algorithm is semantically equivalent to the following code.

\vspace{-0.25cm}
{\scriptsize
\begin{lstlisting}[language=C,escapechar=@]
for (n in 0..batch)
 for (fout in 0..out_features)
  for (y in 0..H-2)
   for (x in 0..W-2)
    for (fin in 0..in_features)
     for (k0 in 0..3)
      for (k1 in 0..3)
       conv[n, fout, y, x] += weigths[fout, fin, y, x] * input[n, fin, y+k0, x+k1];
\end{lstlisting}
}

The next code shows an example of code transformation commands that can be applied to the previous convolution kernel. These commands apply parallelization, loop interchange, tiling, vectorization, and unrolling.

\vspace{-0.25cm}
{\scriptsize
\begin{lstlisting}[language=C,escapechar=@]
// Provide the code transformation commands.
conv.parallelize(n);
conv.interchange(fout, fin);
conv.tile(y, x, 32, 32);
conv.vectorize(fout, 8);
conv.unroll(k0); conv.unroll(k1);
\end{lstlisting}
}

\vspace{-0.25cm}
Currently, in \framework{}, a developer has to provide the previous sequence of code transformations manually. Our goal is to automate finding that sequence. We do this by developing a cost model that predicts the speedup of using a given transformation or any sequence of valid transformations. For example, the model can be used to predict whether combining parallelization, loop interchange, and loop tiling is useful. In addition, the model can be used to choose the right arguments for each one of the previous code transformations (e.g., choose the tile sizes).


\vspace{-0.25cm}
\section{Data Generation}

As training DNNs requires a large data set and only a small number of programs have ever been written in \framework{}, we decided to automatically generate a data set and use it to train the model. We developed a code generator that generates random programs and sequences of code transformations. Each one of these randomly generated programs and code transformations is compiled, executed, and finally, the actual speedup is measured. The speedup is the ratio between the execution time of the original unoptimized program and the optimized one. Each data point in the data set is a triplet of the form (program, a sequence of code transformations, measured speedup).

\vspace{-0.35cm}
\paragraph{Random Code Generation}
A \framework program is a sequence of computations where each computation is an assignment.
There are three common patterns of assignments that appear in \framework programs: (1) simple assignments where the right-hand side is a function of input arrays or array values computed previously; (2) stencils; (3) reductions. The random code generator generates sequences of computations where each computation is a variant (or a combination) of the previous patterns.
Randomly generated programs are correct by construction. A computation consumes either constants, input arrays, or values computed by previous computations. Code transformations are also generated randomly but specific rules are used to guarantee that code transformations are valid (for example, tiling is not applied if the loop extent is smaller than the tile size).

\iflong

The random code generator is designed to generate
programs that are representative of real programs.
The three patterns that the random code generator generates, when combined together, cover all the patterns that we are interested in supporting.
The input data is also generated automatically and the size of the input data is chosen randomly. Since the generated programs are not data dependent (no data dependent conditional or data dependent array access), the actual data values are not important. Only the size of the data is important for training the model.

For a given randomly generated program, the random code generator chooses random data sizes, and then generates 32 random sequences of code transformations.


\vspace{-0.25cm}
\paragraph{Dataset Construction}
The total generated dataset has approximately \DataSetSize programs.
To construct this dataset, we generated $56250$ random algorithms. For each algorithm, we generated $32$ random sequences of code transformations. Therefore we obtained $56250 \times 32$ programs in total. We followed the gold-standard in performance engineering and executed each resulting program 30 times, and retained the median value of the execution times in order to reduce the impact of minor variance in execution times. Since data generation is time consuming, we used a cluster of 16 nodes of multicore CPUs to accelerate data generation. Generating the whole data set took 3 weeks.

\fi

\vspace{-0.25cm}
\section{Program Characterization and Model Architectures}

\begin{figure*}[!h]
\vspace{-0.75cm}
\centering
\includegraphics[width=\linewidth]{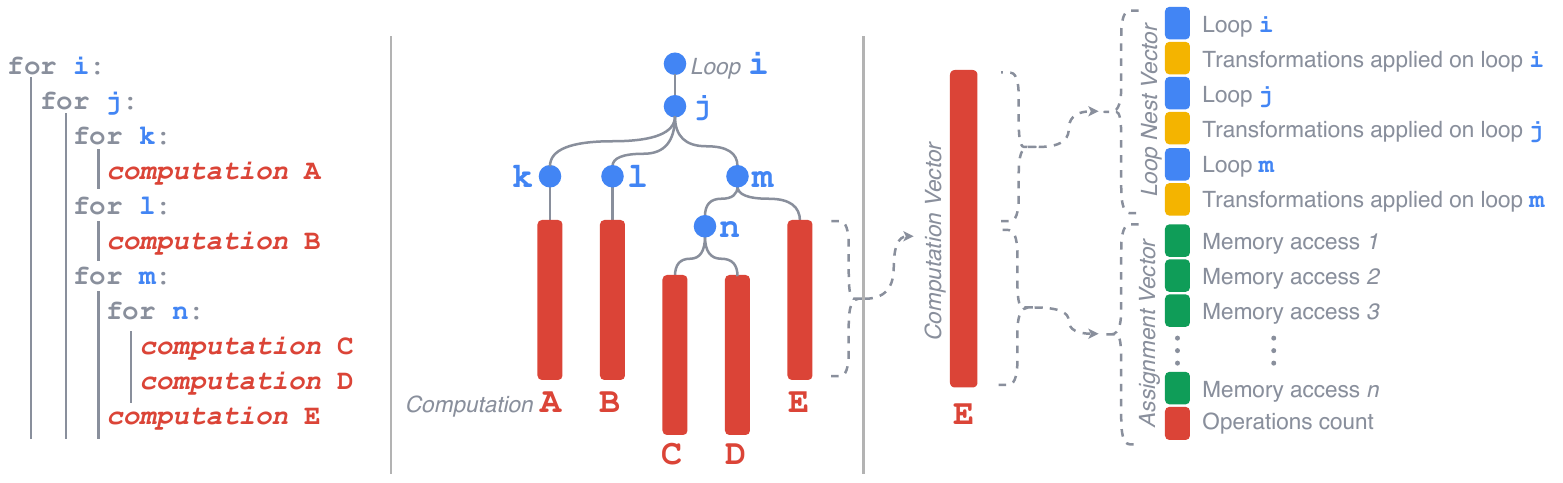} 
\begin{minipage}[t]{.22\linewidth}
\centering
\vspace{-12pt}
\subcaption{Program pseudocode.}\label{fig:egProgam}
\end{minipage}
\begin{minipage}[t]{.34\linewidth}
\centering
\vspace{-12pt}
\subcaption{Program tree representation.}\label{fig:programTree}
\end{minipage}
\begin{minipage}[t]{.40\linewidth}
\centering
\vspace{-12pt}
\subcaption{Computation vector.}\label{fig:compVector}
\end{minipage}
\caption{Our characterization of a typical program.\label{fig:Representation}}
\vspace{-12pt}
\end{figure*}

Our cost model is designed to support programs that can be expressed in \framework. The latter is designed for expressing data parallel algorithms that operate over dense arrays using loop nests and sequences of statements. These algorithms are often found in image processing, deep learning, dense linear algebra, tensor operations, and stencil computations. A formal description of programs supported by \framework can be found in ~\cite{tiramisu,baghdadi2020tiramisu}. Code transformations supported by the proposed model include loop fusion, loop tiling, loop interchange, and loop unrolling which are all challenging. For simpler transformations such as parallelization and vectorization, we use simple heuristics similar to those used by the Halide autoscheduler \cite{halide_12}. These heuristics mainly parallelize the outermost loops and vectorize the innermost loops when a set of conditions are met.

\subsection{Program Characterization}
\vspace{-0.25cm}


Designing complex hand-engineered features is tedious, error-prone, and time-consuming.
Instead of using complex hand-engineered features, we characterize programs by extracting simple high-level information that is stored in a compact variable-size representation.




Our program characterization is based on the AST (Abstract Syntax Tree) representation of programs.
A program is characterized as an ordered tree of \emph{computation vectors} as shown in Figure \ref{fig:programTree}.  
A \emph{computation vector} is a vector that includes three pieces of information:  (1) \emph{loop nest representation}; (2) \emph{assignments representation}; (3) \emph{loop transformation representation}.
\iflong
In the following paragraphs, we describe each of these components, the key features we aim to encode by each one, and how we combine them in a compact way.

\vspace{-0.3cm}
\paragraph{Loop Nest Representation} The extent of each loop level around the computation is stored in the computation vector (the extent of a loop is calculated based on its lower and upper bounds). An example is shown in Figure~\ref{fig:compVector}.
After each loop level extent, for each loop transformation, we insert a boolean tag that represents whether that transformation is applied to that particular loop level. Each transformation is followed by its parameters (when this applies).

\vspace{-0.25cm}
\paragraph{Assignments Representation}
We represent both the left-hand side and the right-hand side of the assignment. To represent the left-hand side, we store the dimensions and the size of each dimension of the buffer used on the left-hand side.
To represent the right-hand side of the assignment (the assignment expression), we store the following information: (1) the memory access pattern (access matrix described later); (2) the ID of each accessed buffer (a number); (3) the count of each arithmetic operation used on the right-hand side (i.e., the number of times each arithmetic operation is used).

We represent the array accesses using an access matrix that stores the coefficients of each array access. This matrix uses exactly the same format used in the polyhedral model to represent array accesses~\cite{polyhedral}. It only supports arrays that have affine array accesses (i.e., array accesses that are affine in the loop iterators). Supporting only affine accesses is not a problem since \framework only supports code that has affine accesses. Supporting code that has data-dependent array accesses or non-affine accesses is not within the scope of this paper.

The access matrix has $k$ rows and $n+1$ columns where $k$ is the number of dimensions of the access buffer and $n$ is the loop depth. Each row in the matrix represents an array dimension. Each array dimension is considered to be a linear combination of the loop iterators. Each loop iterator in the matrix is represented by a column. The coefficient of each loop iterator is stored in the column that corresponds to that loop iterator. The last column in the matrix corresponds to constants.

\vspace{-0.65cm}
\begin{equation*}
M=\begin{bmatrix}
1 & 0 & 0\\
1 & 1 & 0\\
0 & 1 & -2
\end{bmatrix}
\end{equation*}
\vspace{-0.5cm}

For example, the following memory access $ A[ i_{0} ,\ i_{0} +i_{1} ,\ i_{1} -2]$ is represented using the matrix $M$.
In this case, the first dimension of the buffer access is $i_{0}$. It can also be written as $1*i_0 + 0*i_1 + 0$. It is represented by the first row in the matrix using $1~0~0$ where the first column corresponds to iterator $i_0$, the second column corresponds to $i_1$ and the last column corresponds to constants. The second access $i_{0} +i_{1}$ is represented with the second row in the matrix $1~1~0$.
Each \emph{memory access matrix} is succeeded by the identifier of the buffer.


\vspace{-0.25cm}
\paragraph{Loop Transformation Representation} Each loop transformation can be characterized by two pieces of information: transformation type and its parameters.
Since transformations are applied on loop levels, we attach to the representation of each loop level the transformations that are applied to that level (Figure~\ref{fig:compVector}).
The transformations that involve changing the structure of the program (e.g. loop fusion) are directly applied to the \emph{program structure representation} that we will describe next.

\vspace{-0.25cm}
\paragraph{Program Structure Representation} The program is represented as a tree structure where leaves are the \emph{computation vectors} and internal nodes are the loop levels, as shown in Figure~\ref{fig:programTree}. 

\vspace{-0.25cm}
\subsection{Detailed List of Features Composing the \emph{Computation Vector}}
\label{representationAppendix}

\begin{table*}[h]
\footnotesize
\centering
\begin{tabular}{@{}lll@{}}
\toprule
\multirow{7}{*}{\emph{Loop Nest Vector}}   & \multirow{2}{*}{Loop$_1$} & Upper bound, Lower bound, Reduction tag                                                                                 \\
                                    &                        & Fusion tag, Interchange tag, Tilling tag, Tilling factor                                                                          \\ \cmidrule(l){2-3} 
                                    & \multirow{2}{*}{Loop$_2$} & Upper bound, Lower bound, Reduction tag                                                                                        \\
                                    &                        & Fusion tag, Interchange tag, Tilling tag, Tilling factor                                                                          \\ \cmidrule(l){2-3} 
                                    & \vdots                    & \vdots                                                                                                                         \\ \cmidrule(l){2-3} 
                                    & \multirow{2}{*}{Loop$_n$} & Upper bound, Lower bound, Reduction tag                                                                                         \\
                                    &                        & \begin{tabular}[c]{@{}l@{}}Fusion tag, Interchange tag, Tilling tag, Tilling factor,\\  Unroll tag, Unrolling factor.\end{tabular}  \\ \midrule
\multirow{5}{*}{\emph{Assignment Vector}} & Memory access $_1$        & Access matrix,  Buffer ID.                                                                                           \\ \cmidrule(l){2-3} 
                                    & Memory access $_2$        & Access matrix,  Buffer ID.                                                                                                   \\ \cmidrule(l){2-3} 
                                    & \vdots                    &\vdots                                                                                                                            \\ \cmidrule(l){2-3} 
                                    & Memory access $_m$        & Access matrix,  Buffer ID.                                                                                                   \\ \cmidrule(l){2-3} 
                                    & Operations count       & \begin{tabular}[c]{@{}l@{}}Number of Additions, Number of Multiplications,\\ Number of Subtractions, Number of Divisions.\end{tabular}           \\ \bottomrule
\end{tabular}
\vspace{-0.25cm}
\caption{A detailed listing of the features that compose the \emph{Computation Vector}.}
\label{Table:ReprDetails}
\end{table*}
\vspace{-0.25cm}

Table~\ref{Table:ReprDetails} details the full list of features that constitute the \emph{Computation Vector} presented in Figure~\ref{fig:compVector}.
All features are integer values except \emph{Tag} features which are boolean values.
In practice, we set $n=7$ and $m=21$ where $n$ is the maximum length of the loop nests in our dataset and $m$ is the maximum number of terms used in an assignment.
When needed, a zero-padding is added to the \emph{Loop Nest Vector} and \emph{Assignment Vector}.

\vspace{-0.25cm}
\subsection{Hardware Characterization}
\vspace{-0.25cm}

The goal of the paper is not to develop a hardware independent model. The proposed model is specific to a particular CPU. The user can build a model for each CPU they want to target. Building a new model does not require a large effort. It can be done simply by running a script that generates new data and retrains the model.

In other words, we do not include any feature to represent the hardware because the model is specific to one and only one hardware architecture. We also do not extract any hardware-specific feature from the code for the same reason. Designing a model that works for multiple hardware architectures is beyond the scope of this paper.


\fi

\vspace{-0.25cm}
\subsection{Model Architecture}
\vspace{-0.25cm}


\begin{figure*}[!h]
\centering
\includegraphics[width=\linewidth]{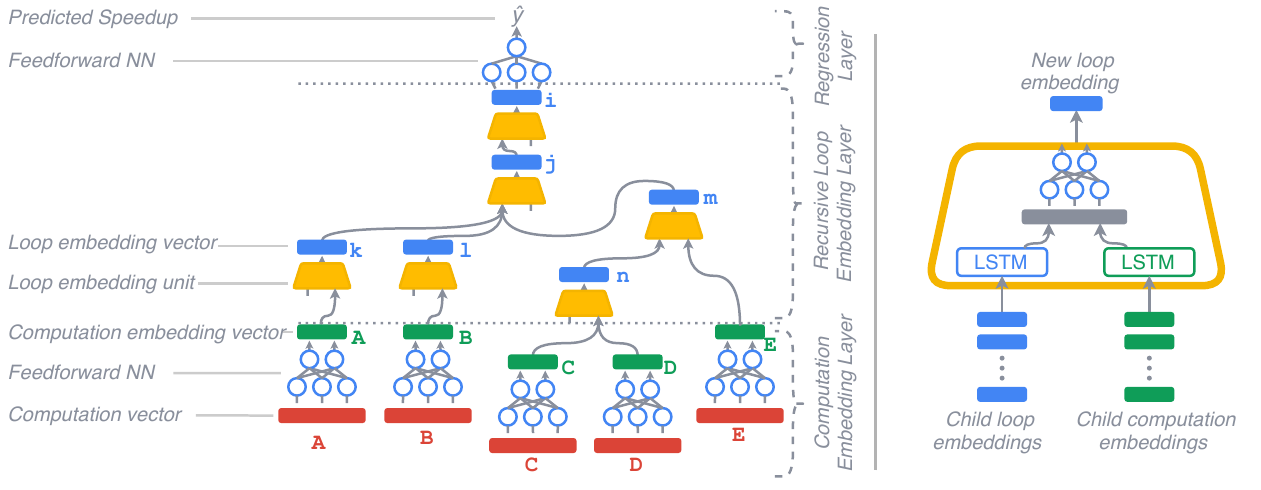} 
\begin{minipage}[t]{0.6\linewidth}
\centering
\vspace{-8pt}
\subcaption{Processing the program presented in Figure~\ref{fig:Representation} through the three layers of the cost-model.}\label{fig:FullModel}
\end{minipage}
\hspace{1cm}
\begin{minipage}[t]{0.2\linewidth}
\centering
\vspace{-8pt}
\subcaption{Loop embedding unit.}\label{fig:LPU}
\end{minipage}
\vspace{-12pt}
\caption{The cost model architecture}
\vspace{-6pt}
\end{figure*}

We model the problem of speedup estimation as a regression problem: given an algorithm and a set of code transformations, our model predicts the speedup expected when applying the suggested code transformations compared to the base program (i.e. without applying code transformations).

%

We design our cost model's architecture to support the variable size and recursive nature of our program characterization by combining Recurrent and Recursive Neural Networks.
Our model's architecture has three layers as shown in Figure~\ref{fig:FullModel}.

\iflong

\vspace{-0.35cm}
\paragraph{Computation Embedding Layer} All \emph{computation vectors} of the program are processed through a feedforward network after \emph{log}-transforming  non-boolean features. 
This \emph{log}-transformation is necessary since these features have a large dynamic range and most of them are expected to be multiplied with other features to compute the final speedup.

\vspace{-0.35cm}
\paragraph{Recursive Loop Embedding Layer} The \emph{computation embeddings} resulting from the previous layer are then processed recursively following the tree structure of the program using the \emph{loop embedding unit}. At a given loop level, the loop embedding unit summarizes the program up to that loop into a \emph{loop embedding}. 
The \emph{loop embedding unit}, as depicted in Figure~\ref{fig:LPU}, is composed of two separate LSTM cells \cite{10.1162/neco.1997.9.8.1735} and a feedforward layer. 
At each loop level, the first LSTM is fed with the embedding vectors of computations that are nested directly in that loop level while the second LSTM is fed with the embedding vectors of the previous loop levels that resulted from the previous \emph{loop embedding units}. The two hidden states of the LSTMs are merged using a feedforward layer into a \emph{loop embedding vector}.

The purpose of this recursive embedding is to selectively incorporate information from each computation respecting its positional relations with the other computations. The output of this layer, the \emph{program embedding vector}, is assumed to contain the needed set of automatically extracted features covering the complete program.


\vspace{-0.35cm}
\paragraph{Regression Layer} The \emph{program embedding vector} produced by the previous layer is finally fed to a shallow feedforward neural network that performs a regression in order to predict the speedup.

We choose MAPE (Mean Absolute Percentage Error) as an objective function to train our model. The model is implemented in PyTorch \cite{NEURIPS2019_9015} and optimized using AdamW \cite{DBLP:journals/corr/abs-1711-05101}. All the implementation details including layer sizes and training policy can be found in appendix \ref{modelDetails}.

\vspace{-0.35cm}
\paragraph{Other Neural Network Models Explored}
We also explored many other alternative architectures for the cost-model, the architecture presented above has the lowest MAPE error on both the test set and benchmarks set. For instance, replacing the \emph{Recursive loop embedding layer} with a simple Recurrent Neural Network that is directly fed with the sequence of \emph{computation embeddings} without taking in consideration the loops hierarchy leads to a relative increase of 1.15x in MAPE of the test set and 1.33x in the benchmarks set compared to the presented architecture. Another straightforward choice of using a simple Feedforwad Neural Network, i.e. totally skipping the \emph{Recursive loop embedding layer} and feeding directly the concatenated \emph{computation embeddings} to the \emph{regression layer}, leads to a relative increase of 1.39x in MAPE of the test set and 1.37x in the benchmarks set compared to the presented model. In addition, this alternative has the considerable limitation of not supporting variable program sizes and supports only programs that contain up to a certain number of computations (we have set the maximum number of computations to 4 when testing this alternative).

\vspace{-0.25cm}
\subsection{Level of Feature Extraction}
\vspace{-0.25cm}

One of the questions that we needed to answer is at which level should the code representation be extracted: directly from the source code or from the transformed code (intermediate representation obtained after applying code transformations)?
Our choice was to extract the representation from the \framework source code for a pragmatic reason. In order for a model that takes transformed code to work, Tiramisu will need to apply the transformations on the program before using the model. Such a step is time-consuming especially because it will be repeated a large number of times (given that the search space is large). Furthermore, transformed code is more complex and therefore is harder to learn from compared to a program and a list of transformations.

\fi

\vspace{-0.25cm}
\section{Search Space Exploration}

Finding the best code transformations is a hard combinatorial optimization problem due to the fact that some of the constraints (e.g., interaction between code transformations), and the objective (the speedup in this case), are hard to represent mathematically using the program's features. 
Thus, the proposed model is used as an objective function estimator to better navigate the search space. 
However, the used search exploration approach should take into account the estimator's margin of error, thus requiring stochasticity in the search space exploration. 

Since the interaction between code transformations is hard to characterize, one of the best ways to model the problem of finding the best code transformations (and their parameters) is to use a tree search. This allows us to use classical tree search algorithms. In this paper, we use Beam Search and MCTS (Monte Carlo Tree Search).

\begin{figure}
{
    \vspace{-0.25cm}
    \centering
    \includegraphics[scale=0.22]{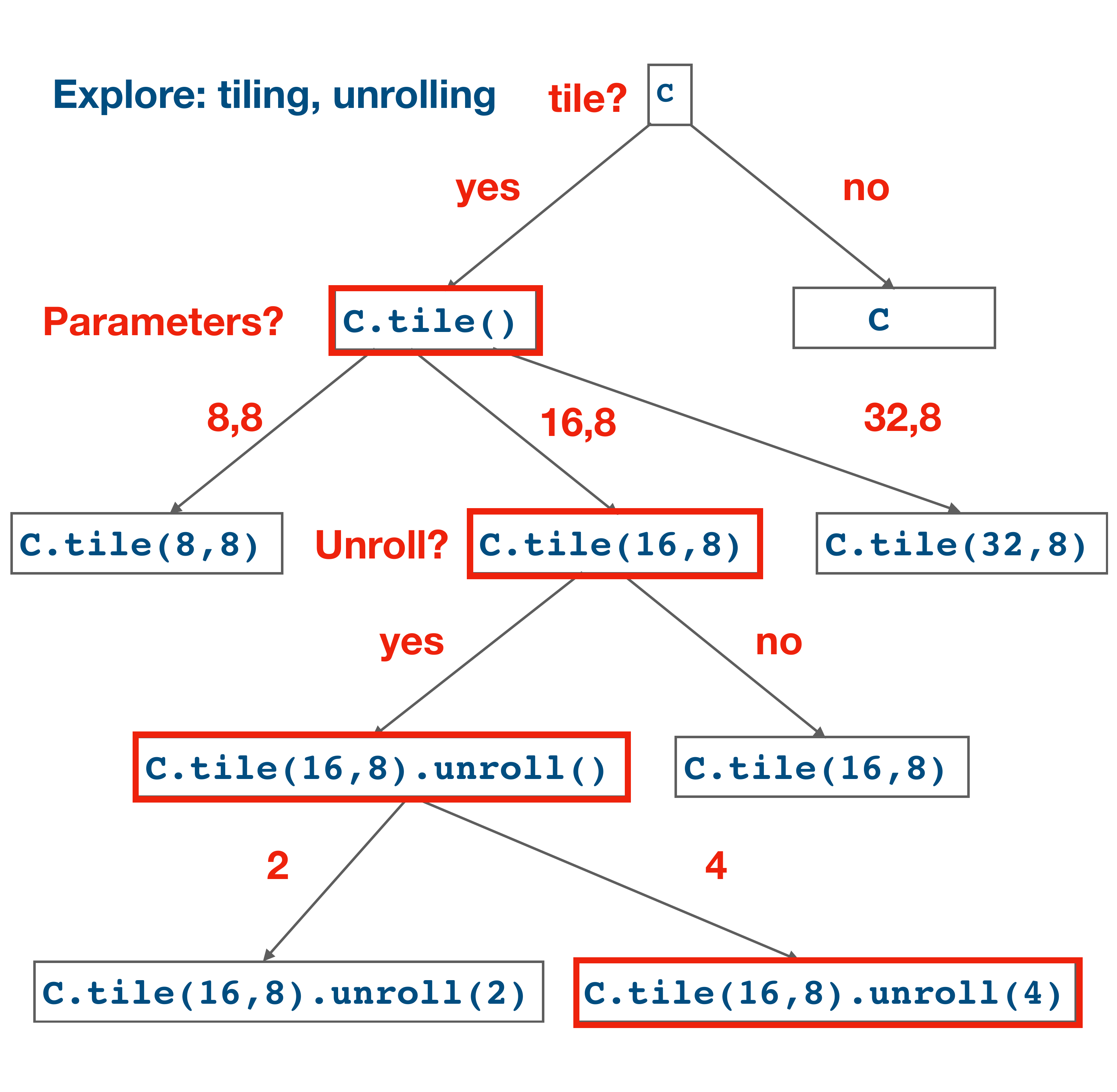}
    \vspace{-0.5cm}
    \caption{Example of the BS Tree for exploring the tiling and unrolling code transformations}
        \vspace{-0.25cm}
    \label{fig:search-space:BSTree}
}
\end{figure}

The Beam Search tree (as shown in Figure \ref{fig:search-space:BSTree}) explores whether to apply a code transformation and which parameters to use for that transformation.
At each node of the tree, an evaluation is conducted using the cost model to assess whether the chosen transformations provide a good speedup.
In Figure \ref{fig:search-space:BSTree}, exploring the tree shows that applying tiling with a tile size of (16, 8) and unrolling with a factor of 4 provides the best sequence of code transformations.

\iflong

MCTS takes advantage of the search tree and takes into account the stochasticity of the model. Particularly, the proposed MCTS combines our model's prediction and an execution of the best evaluated code transformations as a compromise between prediction and execution. 
MCTS first explores the different branches of the tree and selects a number of \emph{promising code transformations}. To this end, the model is used in each node to obtain an estimate of the speedup. MCTS keeps track of a set of the best evaluated code transformations to execute them (the size of the set is a parameter of the approach). 
Once the tree is explored, the set of the best code transformations is executed. 
The advantage of this two-step approach is to accelerate the exploration of the search space using the model and to correct the model's error, if occurred, by executing a limited set of programs and their code transformations. 
The proposed model is thus used to prune the search space and limit the execution to selected code transformations. 



\fi

\vspace{-0.25cm}
\section{Evaluation}
\label{sec:evaluation}
To evaluate our cost model: (1) we measure its accuracy on a test set composed of random programs and compare the predicted and the measured speedups on that data set; (2) we measure the speedups obtained when the model is used to search for code transformations in real-world benchmarks; (3) we compare the accuracy of this model with the accuracy of the model used in Halide~\cite{halide_12}, a state-of-the-art model.

The model evaluation and the data collection are performed on 16 identical multi-core CPU nodes. Each node has a dual-socket, each socket is a 12-core Intel Xeon E5-2680v3 CPU, with 128 GB RAM.
We used 60\% of data for training, 20\% for validation, and 20\% for testing.

\vspace{-0.5cm}
{
\footnotesize{
$$MAPE(y, \hat{y}) = \frac{1}{n}\displaystyle\sum_{i=1}^n\Bigl|\frac{y_i - \hat{y}_i}{y_i}\Bigr|$$
}
}
\vspace{-0.5cm}

\vspace{-0.25cm}
\paragraph{Model Accuracy}
To measure the accuracy of the proposed model, we use MAPE (Mean Absolute Percentage Error), where $y$ and $\hat{y}$ are respectively the measured and the predicted speedups. The MAPE of our cost model on the test set is \ModelMAPE{}\%.

The Pearson correlation coefficient for the proposed model is \ModelPearson, showing that the linear correlation between predicted and measured speedups is strong. In addition, we evaluate the ranking capabilities of the model with the Spearman's rank correlation coefficient, defined as:
$r_s(y, \hat{y}) = r\bigl(rg(y), rg(\hat{y})\bigr)$
where $rg(y)$ converts the speedups to ranks and $r$ is the Pearson correlation coefficient. The Spearman's rank coefficient of our cost model is \ModelSpearman, which shows that the predicted and measured ranks are highly linearly correlated. This property is important when using the model with a search method.

\vspace{-0.25cm}
\paragraph{Comparing Predicted and Measured Speedups}

\begin{figure*}[ht!]
  \vspace{-0.25cm}
  \centering
    \includegraphics[width=\textwidth]{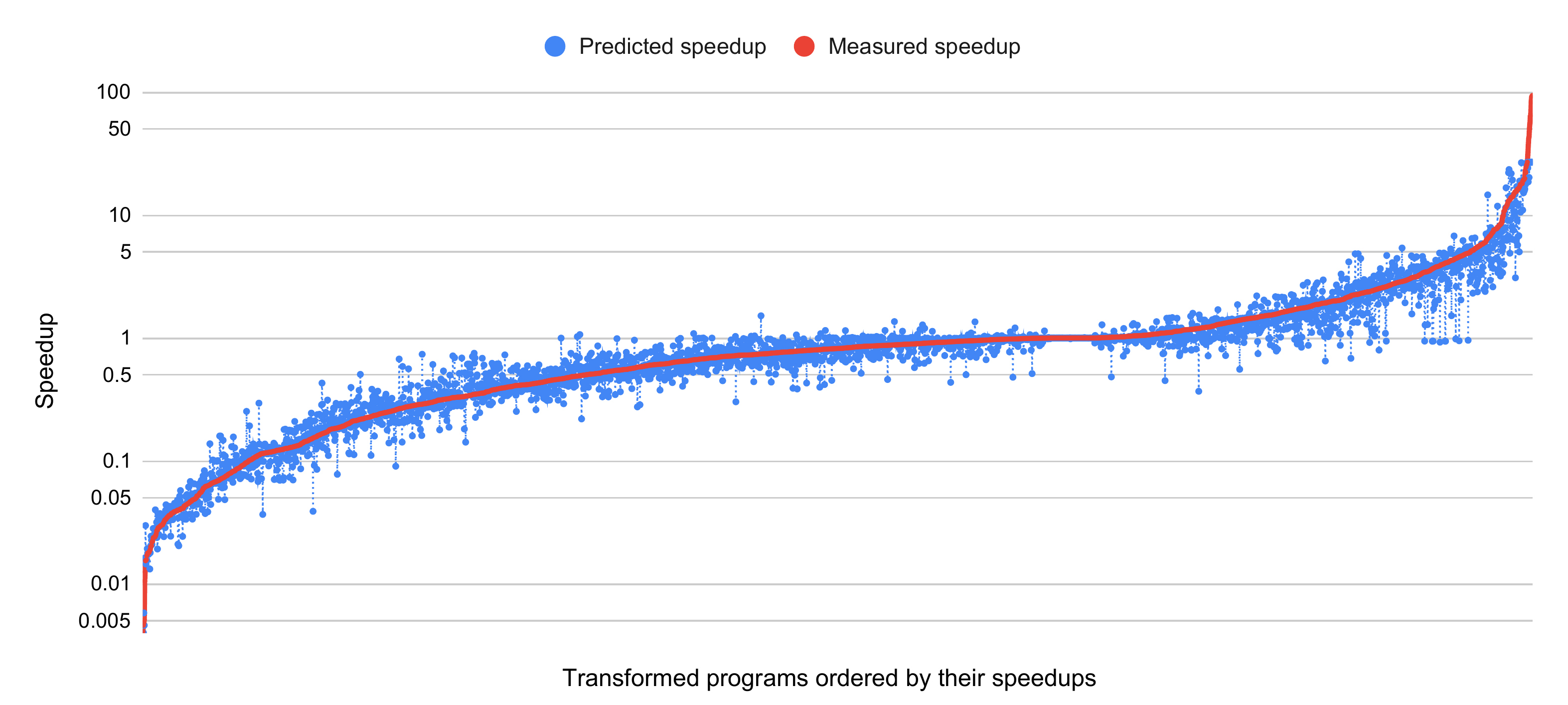}
  \caption{Predicted speedups compared to measured speedups. The speedups are ordered in ascending order.}
  \label{measuredVSspeedup}
  \vspace{-0.25cm}
\end{figure*}

Figure~\ref{measuredVSspeedup} compares the predicted and measured speedups. To simplify visualization, we use a subset of the test set. This subset is composed of 100 random programs, each with 32 random sequences of code transformations (therefore, the total is 3200 transformed programs). The horizontal axis is the list of 3200 programs. These programs are sorted based on their speedups in ascending order to simplify visualization. As the figure shows, the predicted speedups are close to the measured ones.
The error in prediction is lower around the speedup 1 and is higher as the speedup gets further from 1. We will comment more on this behavior later in the section.



\begin{figure}[h!]
\vspace{-0.25cm}
\centering
  \begin{minipage}{0.4\textwidth}
        \includegraphics[width=\textwidth]{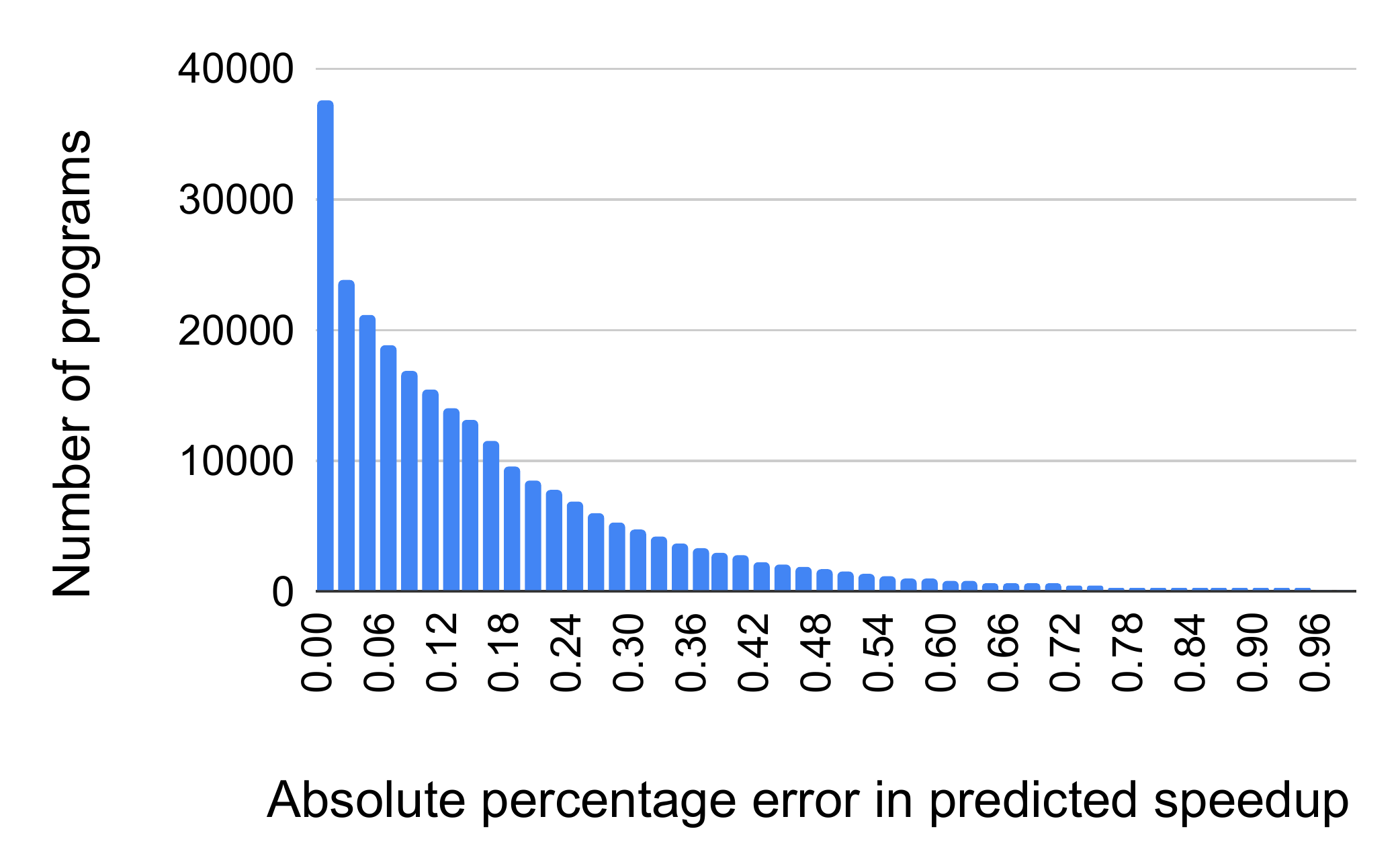}
  \end{minipage}
  \hfill{}
  \begin{minipage}{0.45\textwidth}
        \includegraphics[width=\textwidth]{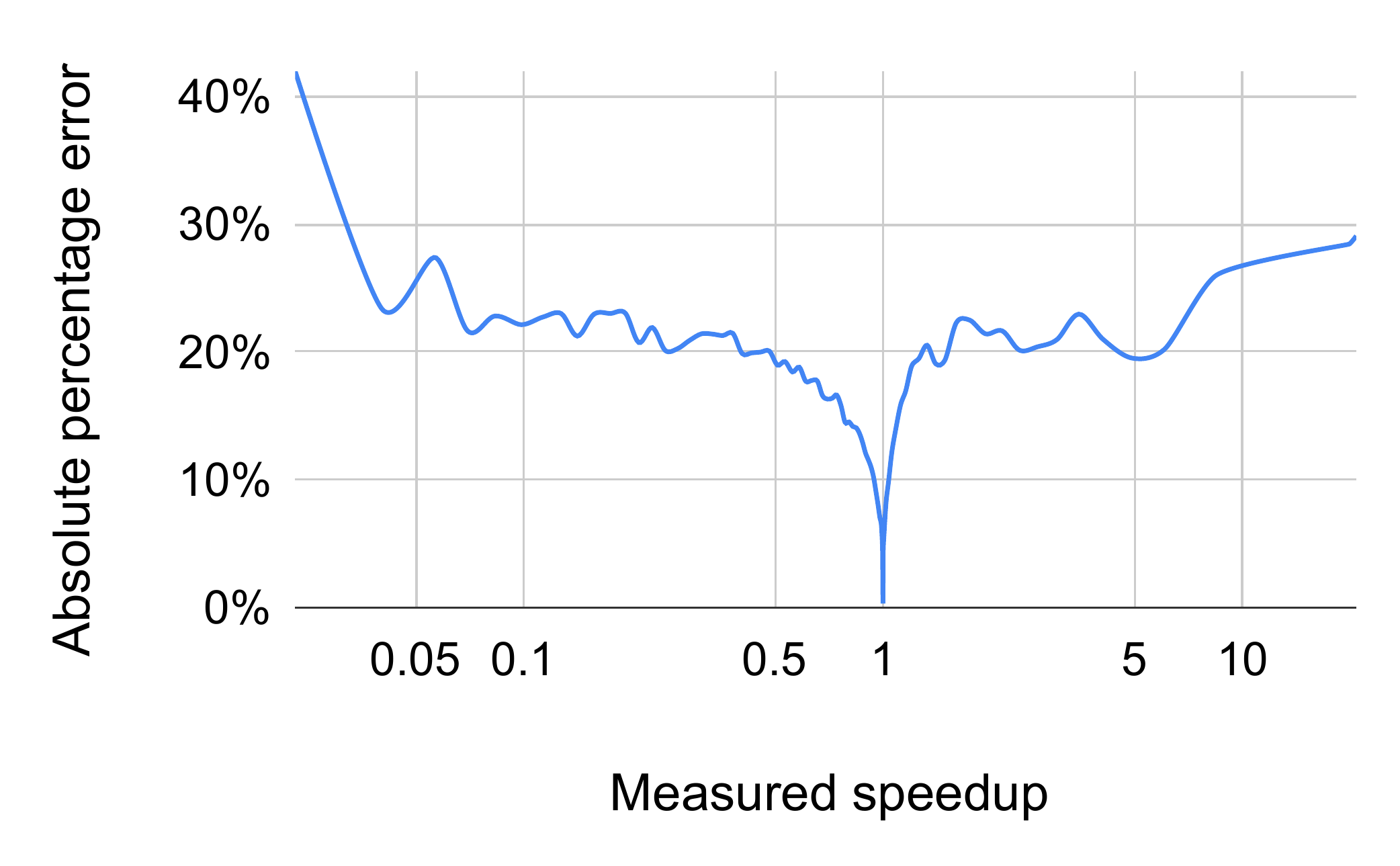}
  \end{minipage}
  \caption{The distribution of error rates for the whole test set. On top, APE is measured for each transformed program, then the histogram of measurements is plotted. On bottom, APE is measured for each transformed program, then the speedups are plotted with their APE.}
  \label{HistogramOfErrors}
  \vspace{-0.65cm}
\end{figure}

Figure~\ref{HistogramOfErrors} investigates the distribution of the model error rates over the whole test set. On top, Absolute Percentage Error (APE) is measured on the code transformations of each program and the results are plotted through a histogram.
On bottom, APE is measured on all data points of the test set and the measured speedups are plotted against their APE. We can see that the error gets smaller as speedups approach 1 and gets higher as speedups get far from 1. Particularly, the error is more significant for speedups below 0.05.
The model is more accurate around speedup 1 because most programs in the training data set have speedups close to 1. Speedups below 0.05 are less frequent.
The next experiment will evaluate whether the accuracy of the model allows  finding the best code transformations when searching the space.



\vspace{-0.25cm}
\paragraph{Search Space Exploration Using the Cost Model}
In this experiment, we evaluate the ability of search approach combined with the cost model to find good code transformation sequences for real-world benchmarks. We use BS and MCTS to explore the search space.
We use a set of real-world benchmarks spanning different areas: image processing, deep learning, linear algebra and stencils. The benchmarks include \emph{box blur} (an image processing filter to blur images), \emph{conv + relu} (two successive neural network layers that benefit from operator fusion), \emph{convolution} (a direct neural network convolution), \emph{cvtcolor} (an image processing filter for converting the colors of an input image from RGB to gray), \emph{doitgen} (a kernel from the multiresolution adaptive numerical scientific simulation~\cite{louis-noel_polybench_2010}), \emph{heat2d} (heat equation over 2D space), \emph{heat3d} (heat equation over 3D space), \emph{jacobi2d} (a jacobi-style stencil computation over 2D data with 5-point stencil pattern), \emph{mvt} (matrix vector multiplication composed with another matrix vector multiplication but with transposed matrix), and \emph{seidel2d} (Gauss-Seidel style stencil computation over 2D data with 9-point stencil pattern). The sizes of the input data for each benchmark is provided in the appendix.

Figure~\ref{BenchmarkSpeedups} shows the best speedups found for each benchmark. The baseline is the original program where the outermost loop is parallelized (no other code transformation is applied). The first column (blue), reports results obtained when beam search is used to explore the search space. This column is considered the reference in our comparison as execution is used to obtain the speedups. In the second and third columns, beam search and MCTS use the cost model to predict speedups. The last column shows the speedups obtained after applying the Halide autoscheduler (Halide automatic optimizer) defined in~\cite{Adams:2019:LOH:3306346.3322967}.

\begin{figure}
    \hspace{-27pt}
    \centering
    \includegraphics[width=1.1\columnwidth]{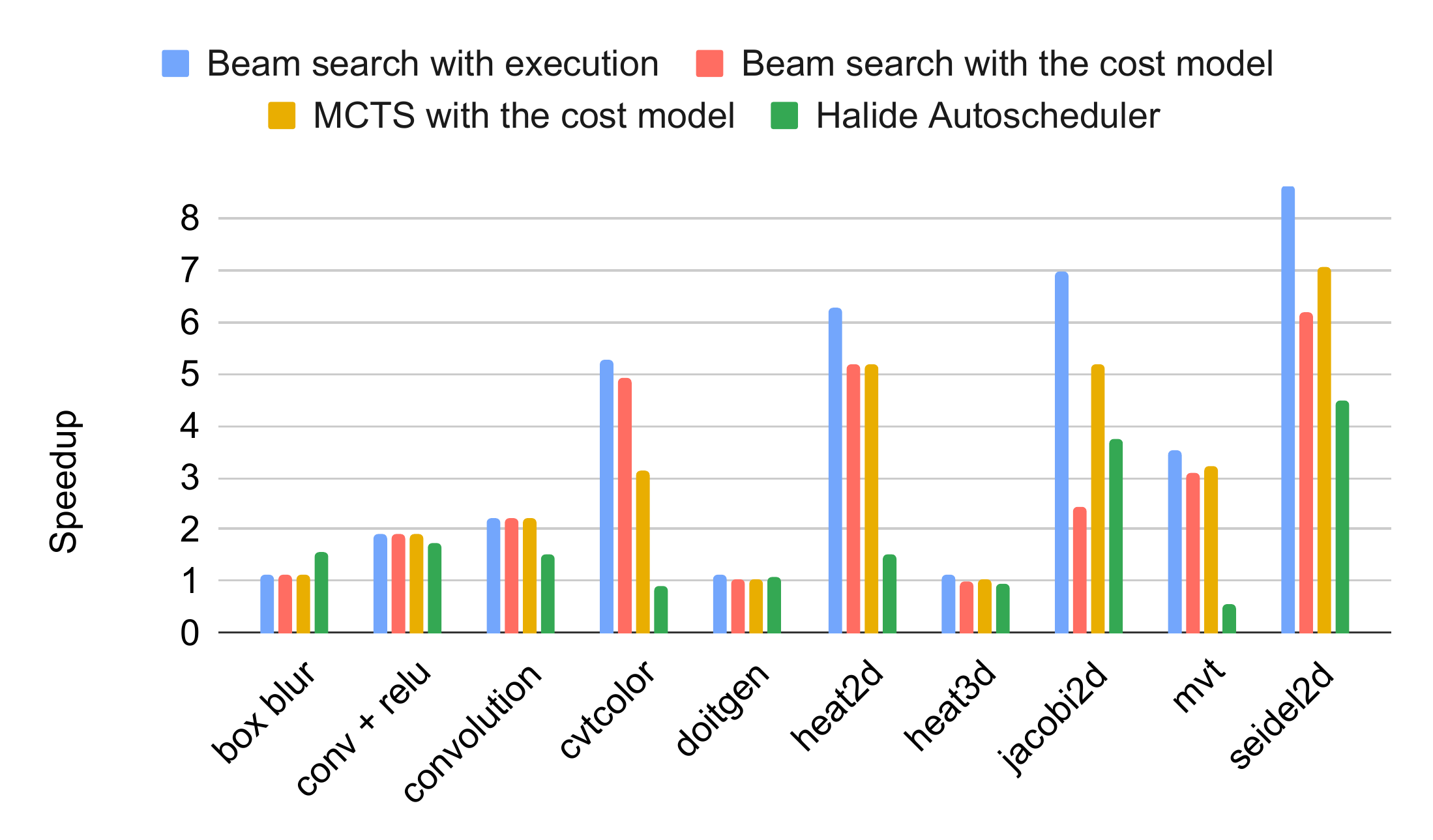}
    \captionof{figure}{Speedups for different benchmarks obtained by exploring the search space.\vspace{-0.5cm}}
    \label{BenchmarkSpeedups}
    
\end{figure}

Beam search (BS) with the cost model is competitive in most benchmarks, but does not find the best code transformations in \emph{heat2d}, \emph{jacobi2d} and \emph{seidel2d}. Beam search with the cost model relies entirely on predictions to make decisions. Bad predictions can thus mislead the search method which is why beam search does not find the best transformations in the previous benchmarks. MCTS has similar performance, except in \emph{jacobi2d} and \emph{seidel2d} where it finds better code transformations, and in \emph{cvtcolor} where the code transformations found are less good. MCTS can find better code transformations in these cases because it copes with model imprecision taking into account its stochasticity. However, since the tree space is explored differently, MCTS might explore different nodes compared to BS and thus have distinguishable results. 

\vspace{-0.25cm}
\paragraph{Comparison with Halide}
In this section, we compare our cost model with the one of Halide~\cite{Adams:2019:LOH:3306346.3322967}, a state-of-the-art cost model and the closest to ours.
In comparison with Halide, \framework finds transformation sequences that are either competitive with those found by Halide or better (except in \emph{box blur}). This is mainly due to miss predictions by the Halide model which lead Halide to use transformations that degrade performance. These wrong predictions happen in particular in benchmarks that are from the area of scientific computing which Halide was not trained to handle (\emph{heat2d}, \emph{jacobi2d}, \emph{mvt} and \emph{seidel2d}). In benchmarks that fall in the categories of deep learning and image processing, which Halide supports well, \framework and Halide have comparable performance.


We also compare the performance of the Halide model with that of \framework on randomly generated programs.
Halide's paper uses $R^2$ as an accuracy metric and uses MSE (Mean Square Error) as a loss function, we thus use the same metric and loss function for comparison. Halide has an $R^2$ of 0.96, whereas \framework has 0.89. Both Halide and \framework have comparable results but Halide uses heavy feature engineering. The main advantage of \framework is that it does not require feature engineering.


\iflong

\paragraph{Tradeoff Between Search Time and Quality of Code Transformations}\label{SearchTimeSchedulesTradeoff}

It is crucial to note that the execution time of a search method is dependent on the time to evaluate code transformations. If one compiles and executes every transformed program to assess the speedup of code transformations, the search time would increase considerably and therefore becomes impractical. The proposed model accelerates the search space exploration, reducing thus the time needed to find the best code transformations.

Table~\ref{SearchTimeVsPerformanceDegradation} illustrates the tradeoff that we make between the time needed to explore the search space, and the performance of the final code transformations found. It compares the gains from reducing the search time and performance degradation due to the use of a model. To be specific, for each benchmark: (1) we compare the time necessary to explore the search space using \emph{Beam Search with Execution} (\emph{BSE}) and using either of \emph{Beam Search with the Cost Model} (\emph{BSM}) on the left table, or \emph{MCTS} on the right table. Comparing these two shows how faster \emph{BSM} and \emph{MCTS} are with regard to \emph{BSE}. (2) we compare the performance of the final code transformations returned by \emph{BSE} and the ones returned by \emph{BSM} or \emph{MCTS}.

The second column of the table represents the speedup in search time (the time taken by \emph{BSE} over the time taken by \emph{BSM} or \emph{MCTS}). The third column is the degradation in performance (execution time) of the code transformations found by \emph{BSM} or \emph{MCTS} compared to the code transformations found by \emph{BSE}. We can see that, on average, searching with \emph{BSM} is 106.5 times faster than searching with \emph{BSE}, while incurring an average decrease of 15\% in the performance of the code transformations found. Moreover, searching with \emph{MCTS} is 11.8 faster on average, with a loss of 12.5\% in performance.

Note that for \emph{jacobi2d} in beam search, and \emph{cvtcolor} in MCTS, the performance degradation is important. This is mainly due to the imprecision of the model for these benchmarks. The model predicts some bad code transformations as being good. This is mainly because a pattern that appears in those benchmarks is not covered enough by the training data. A solution to this problem would be to generate more data that include the missing pattern.

\begin{table*}[h!]
    \centering
    {\scriptsize
    \begin{tabular}{ l >{\centering\arraybackslash}m{1.35cm}  >{\centering\arraybackslash}m{1.35cm} }
    \toprule
    Benchmark & Search time improvement (speedup) & Performance degradation for code transformations \\
    \midrule
    box blur &  \ \ 65x & \ \ 0 \% \\
    conv + relu & \ \ 12x & \ \ 0 \% \\
    convolution & \ \ \ \ 9x & \ \ 0 \% \\
    cvtcolor & \ \ 65x & \ \ 7 \% \\
    doitgen & 114x & \ \ 7 \% \\
    heat2d & 117x & 18 \% \\
    heat3d & 351x & 13 \% \\
    jacobi2d & 122x & 65 \% \\
    mvt & \ \ 83x & 12 \% \\
    seidel2d & 127x & 28 \% \\
    \midrule
    Average & 106.5x & 15 \% \\
    \bottomrule
    \end{tabular}} \hspace{12pt} {\scriptsize
    \begin{tabular}{l >{\centering\arraybackslash}m{1.35cm} >{\centering\arraybackslash}m{1.35cm}}
    \toprule
    Benchmark & Search time improvement (speedup) & Performance degradation for code transformations \\
    \midrule
    box blur & 19x & \ \ 0 \% \\
    conv + relu & 14x & \ \ 0 \% \\
    convolution & 11x & \ \ 0 \% \\
    cvtcolor & \ \ 3x & 41 \% \\
    doitgen & 12x & \ \ 5 \% \\
    heat2d & 17x & 17 \% \\
    heat3d & 28x & \ \ 9 \% \\
    jacobi2d & \ \ 4x & 26 \% \\
    mvt & \ \ 5x & \ \ 9 \% \\
    seidel2d & \ \ 5x & 18 \% \\
    \midrule
    Average & 11.8x & 12.5 \% \\
    \bottomrule
    \end{tabular}}
    
    \vspace{-0.15cm}
    \caption{Search time improvement compared to performance degradation. On the left, the results for beam search with the cost model are shown, and on the right, the results for MCTS.}
    \vspace{-0.5cm}
    \label{SearchTimeVsPerformanceDegradation}
\end{table*}

\vspace{-0.35cm}
\paragraph{More Detailed Evaluation}


\begin{figure}[!h]
\centering
    \includegraphics[width=0.49\columnwidth]{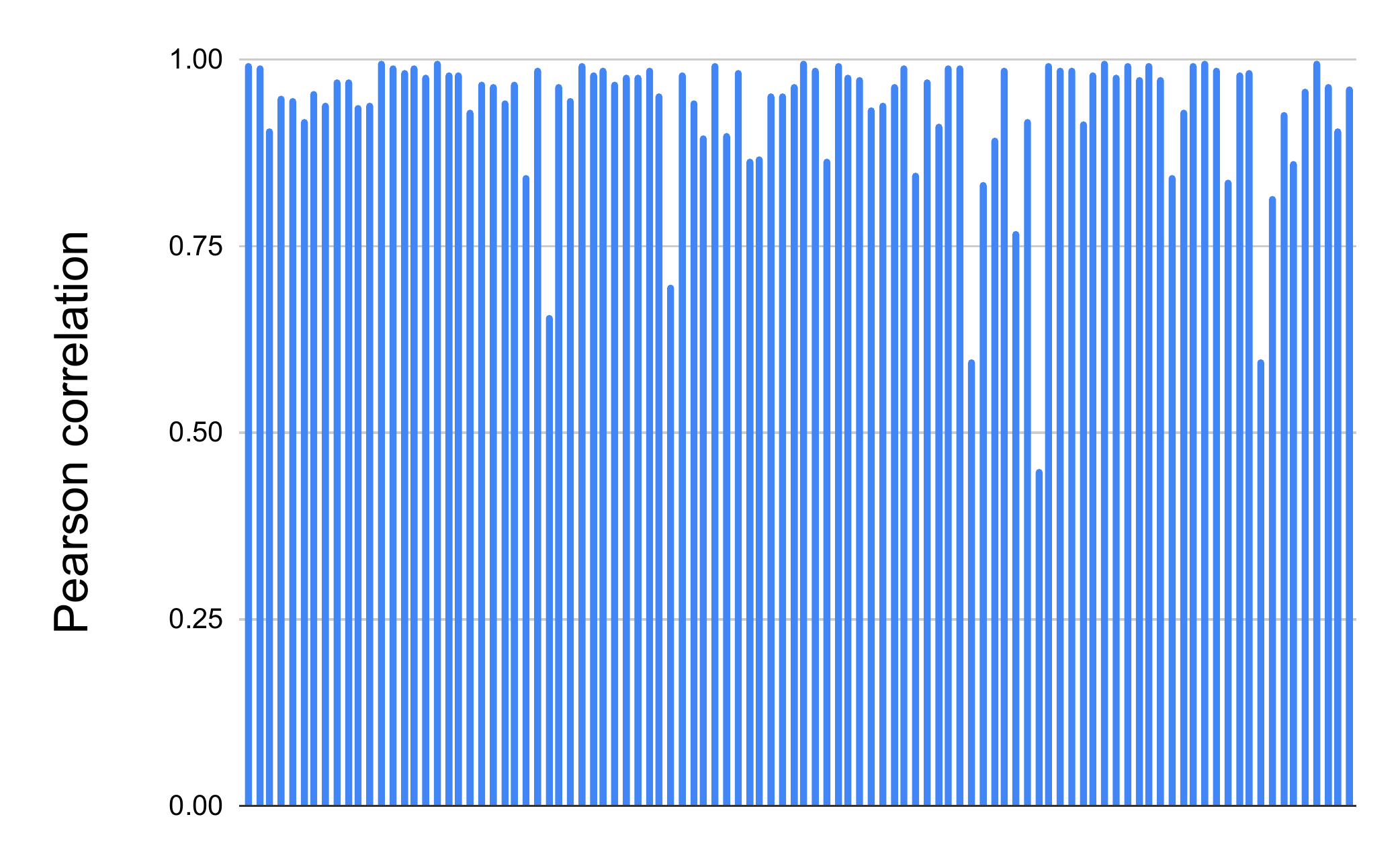}
    \includegraphics[width=0.49\columnwidth]{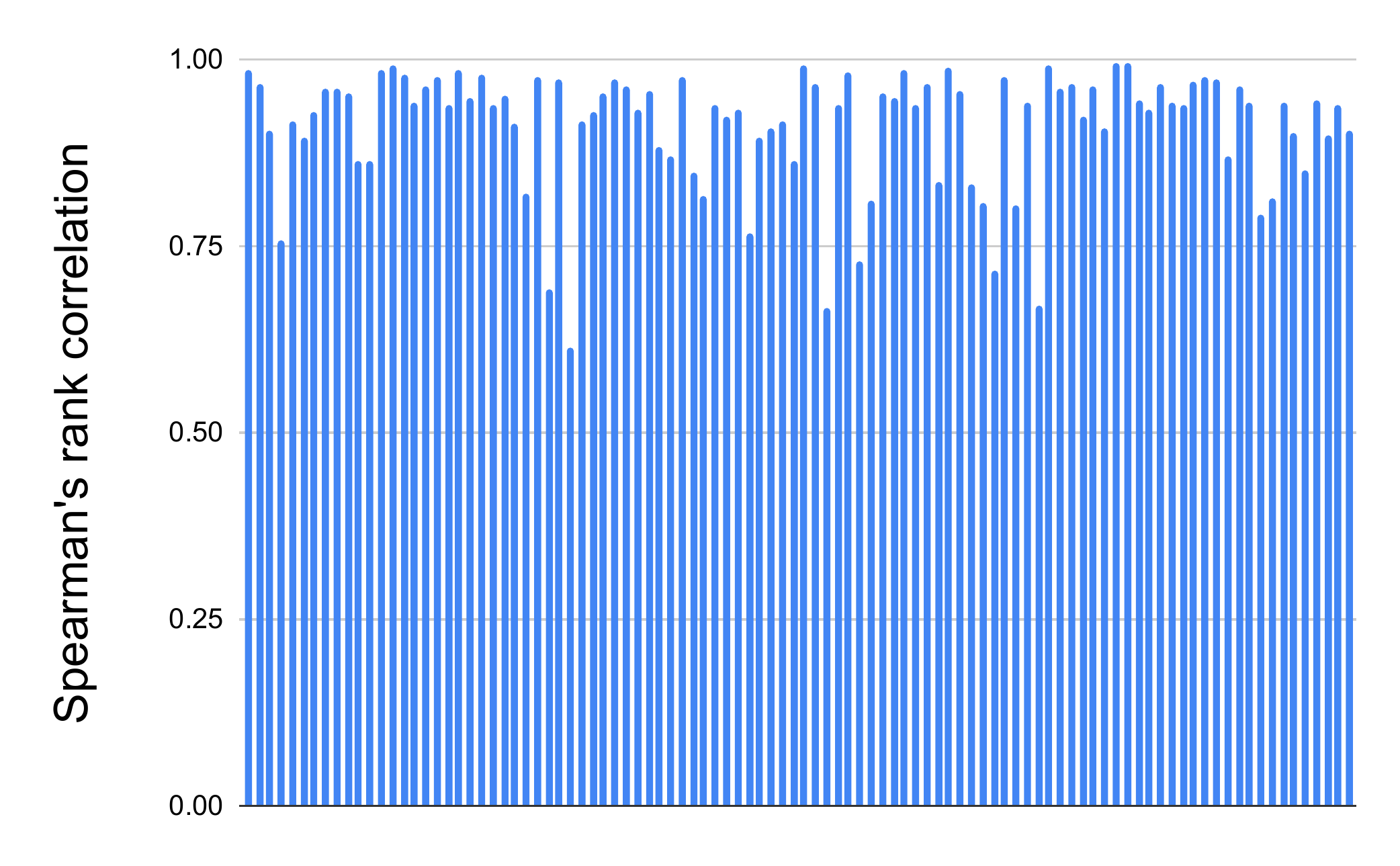}
  \caption{Correlation between predicted and measured speedups for the data points in the \emph{small} test set. Pearson correlation is on the left and Spearman's correlation is on the right. Each column represents the coefficient measured on $32$ random code transformations for a single program.}
  \label{Spearman}
  \vspace{-0.25cm}
\end{figure}
Figure~\ref{Spearman} gives more insights on the behavior of our model. The Pearson and Spearman's coefficients are measured on a subset of the test set between the code transformations of each program. To simplify visualization, we use the same subset of transformations as the experiment of Figure~\ref{measuredVSspeedup}. In most cases, the coefficients are close to 1, highlighting a strong linear correlation between the predicted and measured speedups (Pearson correlation), and between the predicted and measured ranks (Spearman's correlation). Particularly, the correlation between ranks is of most importance for search methods as these methods rely on the ranks of the explored points instead of their actual evaluations.

\vspace{-0.35cm}
\section{Related Work}

Several machine learning based cost models have been developed for automatic code optimization. MILEPOST GCC~\cite{fursin:inria-00294704} uses a 1-nearest-neighbor model that takes manually engineered features and predicts the best combinations of compiler flags for GCC (GNU Compiler Collection).
Ithemal~\cite{ithemal} uses an LSTM based model to predict the throughput of basic blocks of the control-flow graph (assembly-level code).
Both of MILEPOST GCC and Ithemal do not support high-level loop transformations though. Modeling loop transformations is in particular challenging as it requires the ability to model the loop structure. This implies the ability to model cycles in the control-flow graph of the program, which is not trivial. In addition, Ithemal does not model memory accesses to different memory hierarchies (it only models accesses to cache). Unlike both systems, our proposed model supports loop transformations and takes into consideration different memory hierarchy levels.

Other related work includes a model proposed by Rahman et al.~\cite{rahman2010neural}. This model uses a feedforward neural network to predict the execution time of programs after applying the tiling code transformation. Another model proposed by Magni et al.~\cite{10.1145/2628071.2628087} uses a feedforward neural network in a cascade fashion to predict the best thread-coarsening factor. These two methods rely on hand-engineered features though. They are also limited to a single code transformation.


Halide~\cite{Adams:2019:LOH:3306346.3322967} proposes a more comprehensive method to find efficient code transformations. It combines beam search with a feedforward neural network that predicts the execution time of programs from a set of manually-engineered features. It uses 54 heavily engineered features to perform its predictions.
To the best of our knowledge, both of Tiramisu and Halide consider the same code transformations in their search spaces.
In the same context, AutoTVM~\cite{chen2018learning} uses a deep learning model to search for code transformation parameters (TVM compares two models, a TreeGRU and a Gradient Boosted Tree). The TVM models are used with simulated annealing to search the space.
The authors only demonstrate the search for transformation parameters though (tile size, unrolling factor, ...). They do not demonstrate search for loop transformations since the transformations themselves are provided by the user.
In a different fashion, DeepTune \cite{cummins2017end} proposes a neural network consisting of embedding layers and LSTM cells to predict whether an OpenCL kernel should be mapped to CPU or GPU. DeepTune also proposes another model to predict whether thread coarsening (a code transformation for GPUs) should be used. The DeepTune models are classification models though. This means that they can classify whether a given transformation is beneficial, but they are not designed to rank different sequences of code transformations. Therefore, they are not ideal for use when searching a large space of code transformations where many sequences need to be ranked and the best one selected.

The goal of this paper is not to propose a cost model for general purpose compilers. The goal is also not to propose a cost model for general purpose transformations.
In a way similar to related work, this paper mainly focuses on a domain specific compiler and on loop transformations.




Many polyhedral compilers including Pluto~\cite{bondhugula_practical_2008}, PENCIL~\cite{pencil_paper,DBLP:journals/corr/abs-1302-5586}, LLVM Poly~\cite{polly}, and Tensor Comprehensions~\cite{Vasilache2018TensorCF} formalize the problem of automatic code optimization as an integer linear program (ILP). The objective of this ILP is to minimize the distance between producer and consumer statements.
The resulting problem can be solved exactly, but the implicit cost model does not capture all the complexity of the hardware architecture and transformation interactions~\cite{tiramisu}. This leads to suboptimal solutions~\cite{tiramisu,pencil_paper,Baghdadi2013}. Making the objective function more comprehensive makes the problem non-linear and thus it becomes intractable.

\fi


\vspace{-0.35cm}
\section{Conclusion}

This paper presents a novel cost model for predicting speedups. This cost model is a \emph{regression} cost model that operates on \emph{full programs} and \emph{does not rely on extracting complex features}. It is not limited to transformation parameters but also includes code transformations.
We develop a random code generator to generate the training data and release the generator publicly.
We evaluated the proposed model and show that it had a low error rate of \ModelMAPE{}\% MAPE. We integrate this model in a search space method and show that the integrated approach enables \framework to automatically find sequences of code transformations that are competitive with state of the art compilers.




\bibliography{bib}

\begin{thebibliography}{33}
\providecommand{\natexlab}[1]{#1}
\providecommand{\url}[1]{\texttt{#1}}
\expandafter\ifx\csname urlstyle\endcsname\relax
  \providecommand{\doi}[1]{doi: #1}\else
  \providecommand{\doi}{doi: \begingroup \urlstyle{rm}\Url}\fi

\bibitem[Adams et~al.(2019)Adams, Ma, Anderson, Baghdadi, Li, Gharbi, Steiner,
  Johnson, Fatahalian, Durand, and
  Ragan-Kelley]{Adams:2019:LOH:3306346.3322967}
Adams, A., Ma, K., Anderson, L., Baghdadi, R., Li, T.-M., Gharbi, M., Steiner,
  B., Johnson, S., Fatahalian, K., Durand, F., and Ragan-Kelley, J.
\newblock Learning to optimize halide with tree search and random programs.
\newblock \emph{ACM Trans. Graph.}, 38\penalty0 (4):\penalty0 121:1--121:12,
  July 2019.
\newblock ISSN 0730-0301.
\newblock \doi{10.1145/3306346.3322967}.
\newblock URL \url{http://doi.acm.org/10.1145/3306346.3322967}.

\bibitem[Bachir et~al.(2013)Bachir, Brault, Gregg, Cohen,
  et~al.]{bachir2013minimal}
Bachir, M., Brault, F., Gregg, D., Cohen, A., et~al.
\newblock Minimal unroll factor for code generation of software pipelining.
\newblock \emph{International Journal of Parallel Programming}, 41\penalty0
  (1):\penalty0 1--58, 2013.

\bibitem[Baghdadi et~al.(2013{\natexlab{a}})Baghdadi, Cohen, Guelton,
  Verdoolaege, Inoue, Grosser, Kouveli, Kravets, Lokhmotov, Nugteren, Waters,
  and Donaldson]{DBLP:journals/corr/abs-1302-5586}
Baghdadi, R., Cohen, A., Guelton, S., Verdoolaege, S., Inoue, J., Grosser, T.,
  Kouveli, G., Kravets, A., Lokhmotov, A., Nugteren, C., Waters, F., and
  Donaldson, A.~F.
\newblock {PENCIL:} towards a platform-neutral compute intermediate language
  for dsls.
\newblock \emph{CoRR}, abs/1302.5586, 2013{\natexlab{a}}.
\newblock URL \url{http://arxiv.org/abs/1302.5586}.

\bibitem[Baghdadi et~al.(2013{\natexlab{b}})Baghdadi, Cohen, Verdoolaege, and
  Trifunovic]{Baghdadi2013}
Baghdadi, R., Cohen, A., Verdoolaege, S., and Trifunovic, K.
\newblock Improved loop tiling based on the removal of spurious false
  dependences.
\newblock \emph{TACO}, 9\penalty0 (4):\penalty0 52, 2013{\natexlab{b}}.

\bibitem[Baghdadi et~al.(2015{\natexlab{a}})Baghdadi, Beaugnon, Cohen, Grosser,
  Kruse, Reddy, Verdoolaege, Betts, Donaldson, Ketema, Absar, Haastregt,
  Kravets, Lokhmotov, David, and Hajiyev]{pencil_paper}
Baghdadi, R., Beaugnon, U., Cohen, A., Grosser, T., Kruse, M., Reddy, C.,
  Verdoolaege, S., Betts, A., Donaldson, A.~F., Ketema, J., Absar, J.,
  Haastregt, S.~v., Kravets, A., Lokhmotov, A., David, R., and Hajiyev, E.
\newblock Pencil: A platform-neutral compute intermediate language for
  accelerator programming.
\newblock In \emph{Proceedings of the 2015 International Conference on Parallel
  Architecture and Compilation (PACT)}, PACT '15, pp.\  138--149, Washington,
  DC, USA, 2015{\natexlab{a}}. IEEE Computer Society.
\newblock ISBN 978-1-4673-9524-3.
\newblock \doi{10.1109/PACT.2015.17}.
\newblock URL \url{http://dx.doi.org/10.1109/PACT.2015.17}.

\bibitem[Baghdadi et~al.(2015{\natexlab{b}})Baghdadi, Cohen, Grosser,
  Verdoolaege, Lokhmotov, Absar, van Haastregt, Kravets, and Donaldson]{pencil}
Baghdadi, R., Cohen, A., Grosser, T., Verdoolaege, S., Lokhmotov, A., Absar,
  J., van Haastregt, S., Kravets, A., and Donaldson, A.~F.
\newblock {PENCIL} language specification.
\newblock Research Rep. RR-8706, {INRIA}, 2015{\natexlab{b}}.
\newblock URL \url{https://hal.inria.fr/hal-01154812}.

\bibitem[Baghdadi et~al.(2019)Baghdadi, Ray, Romdhane, Del~Sozzo, Akkas, Zhang,
  Suriana, Kamil, and Amarasinghe]{tiramisu}
Baghdadi, R., Ray, J., Romdhane, M.~B., Del~Sozzo, E., Akkas, A., Zhang, Y.,
  Suriana, P., Kamil, S., and Amarasinghe, S.
\newblock Tiramisu: A polyhedral compiler for expressing fast and portable
  code.
\newblock In \emph{Proceedings of the 2019 IEEE/ACM International Symposium on
  Code Generation and Optimization}, CGO 2019, pp.\  193--205, Piscataway, NJ,
  USA, 2019. IEEE Press.
\newblock ISBN 978-1-7281-1436-1.
\newblock URL \url{http://dl.acm.org/citation.cfm?id=3314872.3314896}.

\bibitem[Baghdadi et~al.(2020)Baghdadi, Debbagh, Abdous, Benhamida, Renda,
  Frankle, Carbin, and Amarasinghe]{baghdadi2020tiramisu}
Baghdadi, R., Debbagh, A.~N., Abdous, K., Benhamida, F.~Z., Renda, A., Frankle,
  J.~E., Carbin, M., and Amarasinghe, S.
\newblock Tiramisu: A polyhedral compiler for dense and sparse deep learning,
  2020.

\bibitem[Bondhugula et~al.(2008)Bondhugula, Hartono, Ramanujam, and
  Sadayappan]{bondhugula_practical_2008}
Bondhugula, U., Hartono, A., Ramanujam, J., and Sadayappan, P.
\newblock A practical automatic polyhedral parallelizer and locality optimizer.
\newblock In \emph{PLDI}, pp.\  101--113, 2008.

\bibitem[Chen et~al.(2018)Chen, Zheng, Yan, Jiang, Moreau, Ceze, Guestrin, and
  Krishnamurthy]{chen2018learning}
Chen, T., Zheng, L., Yan, E., Jiang, Z., Moreau, T., Ceze, L., Guestrin, C.,
  and Krishnamurthy, A.
\newblock Learning to optimize tensor programs.
\newblock In \emph{Advances in Neural Information Processing Systems}, pp.\
  3389--3400, 2018.

\bibitem[Cummins et~al.(2017)Cummins, Petoumenos, Wang, and
  Leather]{cummins2017end}
Cummins, C., Petoumenos, P., Wang, Z., and Leather, H.
\newblock End-to-end deep learning of optimization heuristics.
\newblock In \emph{2017 26th International Conference on Parallel Architectures
  and Compilation Techniques (PACT)}, pp.\  219--232. IEEE, 2017.

\bibitem[Feautrier(1988)]{feautrier_array_1988}
Feautrier, P.
\newblock Array expansion.
\newblock In \emph{Proceedings of the 2nd international conference on
  Supercomputing}, pp.\  429--441, St. Malo, France, 1988. {ACM}.
\newblock ISBN 0-89791-272-1.
\newblock \doi{10.1145/55364.55406}.
\newblock URL \url{http://portal.acm.org/citation.cfm?id=55406}.

\bibitem[Fursin et~al.(2008)Fursin, Miranda, Temam, Namolaru, Yom-Tov, Zaks,
  Mendelson, Bonilla, Thomson, Leather, Williams, O'Boyle, Barnard, Ashton,
  Courtois, and Bodin]{fursin:inria-00294704}
Fursin, G., Miranda, C., Temam, O., Namolaru, M., Yom-Tov, E., Zaks, A.,
  Mendelson, B., Bonilla, E., Thomson, J., Leather, H., Williams, C., O'Boyle,
  M., Barnard, P., Ashton, E., Courtois, E., and Bodin, F.
\newblock {MILEPOST GCC: machine learning based research compiler}.
\newblock In \emph{{GCC Summit}}, Ottawa, Canada, June 2008.
\newblock URL \url{https://hal.inria.fr/inria-00294704}.

\bibitem[Glorot \& Bengio(2010)Glorot and Bengio]{glorotunderstanding}
Glorot, X. and Bengio, Y.
\newblock Understanding the difficulty of training deep feedforward neural
  networks.
\newblock \emph{Journal of Machine Learning Research - Proceedings Track},
  9:\penalty0 249--256, 01 2010.

\bibitem[Grosser et~al.(2012)Grosser, Groslinger, and Lengauer]{polly}
Grosser, T., Groslinger, A., and Lengauer, C.
\newblock Polly - performing polyhedral optimizations on a low-level
  intermediate representation.
\newblock \emph{Parallel Processing Letters}, 22\penalty0 (4), 2012.
\newblock URL
  \url{http://dblp.uni-trier.de/db/journals/ppl/ppl22.html#GrosserGL12}.

\bibitem[Grosser et~al.(2014)Grosser, Cohen, Holewinski, Sadayappan, and
  Verdoolaege]{tobias_hexagonal_cgo13}
Grosser, T., Cohen, A., Holewinski, J., Sadayappan, P., and Verdoolaege, S.
\newblock Hybrid hexagonal/classical tiling for gpus.
\newblock In \emph{Proceedings of Annual IEEE/ACM International Symposium on
  Code Generation and Optimization}, CGO '14, pp.\  66:66--66:75, New York, NY,
  USA, 2014. ACM.

\bibitem[Hochreiter \& Schmidhuber(1997)Hochreiter and
  Schmidhuber]{10.1162/neco.1997.9.8.1735}
Hochreiter, S. and Schmidhuber, J.
\newblock Long short-term memory.
\newblock \emph{Neural Comput.}, 9\penalty0 (8):\penalty0 1735–1780, November
  1997.
\newblock ISSN 0899-7667.
\newblock \doi{10.1162/neco.1997.9.8.1735}.
\newblock URL \url{https://doi.org/10.1162/neco.1997.9.8.1735}.

\bibitem[Lefebvre \& Feautrier(1998)Lefebvre and
  Feautrier]{lefebvre_automatic_1998}
Lefebvre, V. and Feautrier, P.
\newblock Automatic storage management for parallel programs.
\newblock \emph{Parallel Computing}, 24:\penalty0 649--671, 1998.
\newblock ISSN 01678191.
\newblock \doi{10.1016/S0167-8191(98)00029-5}.

\bibitem[Loshchilov \& Hutter(2017)Loshchilov and
  Hutter]{DBLP:journals/corr/abs-1711-05101}
Loshchilov, I. and Hutter, F.
\newblock Fixing weight decay regularization in adam.
\newblock \emph{CoRR}, abs/1711.05101, 2017.
\newblock URL \url{http://arxiv.org/abs/1711.05101}.

\bibitem[{Louis-Noel}(2010)]{louis-noel_polybench_2010}
{Louis-Noel}, P.
\newblock {PolyBench} suite.
\newblock
  http://www.cse.ohio-state.edu/{\textasciitilde}pouchet/software/polybench/,
  2010.
\newblock URL \url{http://www.cse.ohio-state.edu/~pouchet/software/polybench/}.

\bibitem[Magni et~al.(2014)Magni, Dubach, and
  O’Boyle]{10.1145/2628071.2628087}
Magni, A., Dubach, C., and O’Boyle, M.
\newblock Automatic optimization of thread-coarsening for graphics processors.
\newblock In \emph{Proceedings of the 23rd International Conference on Parallel
  Architectures and Compilation}, PACT ’14, pp.\  455–466, New York, NY,
  USA, 2014. Association for Computing Machinery.
\newblock ISBN 9781450328098.
\newblock \doi{10.1145/2628071.2628087}.
\newblock URL \url{https://doi.org/10.1145/2628071.2628087}.

\bibitem[Mendis et~al.(2018)Mendis, Amarasinghe, and Carbin]{ithemal}
Mendis, C., Amarasinghe, S.~P., and Carbin, M.
\newblock Ithemal: Accurate, portable and fast basic block throughput
  estimation using deep neural networks.
\newblock \emph{CoRR}, abs/1808.07412, 2018.
\newblock URL \url{http://arxiv.org/abs/1808.07412}.

\bibitem[Paszke et~al.(2019)Paszke, Gross, Massa, Lerer, Bradbury, Chanan,
  Killeen, Lin, Gimelshein, Antiga, Desmaison, Kopf, Yang, DeVito, Raison,
  Tejani, Chilamkurthy, Steiner, Fang, Bai, and Chintala]{NEURIPS2019_9015}
Paszke, A., Gross, S., Massa, F., Lerer, A., Bradbury, J., Chanan, G., Killeen,
  T., Lin, Z., Gimelshein, N., Antiga, L., Desmaison, A., Kopf, A., Yang, E.,
  DeVito, Z., Raison, M., Tejani, A., Chilamkurthy, S., Steiner, B., Fang, L.,
  Bai, J., and Chintala, S.
\newblock Pytorch: An imperative style, high-performance deep learning library.
\newblock In Wallach, H., Larochelle, H., Beygelzimer, A., d'Alch\'{e} Buc, F.,
  Fox, E., and Garnett, R. (eds.), \emph{Advances in Neural Information
  Processing Systems 32}, pp.\  8024--8035. Curran Associates, Inc., 2019.

\bibitem[Paul \& Christian(2011)Paul and Christian]{polyhedral}
Paul, F. and Christian, L.
\newblock The polyhedron model.
\newblock In Padua, D. (ed.), \emph{Encyclopedia of Parallel Computing}, pp.\
  1581, 1592. Springer, 2011.

\bibitem[Quiller\'e \& Rajopadhye(2000)Quiller\'e and Rajopadhye]{Qui00}
Quiller\'e, F. and Rajopadhye, S.
\newblock Optimizing memory usage in the polyhedral model.
\newblock \emph{ACM Trans. on Programming Languages and Systems}, 22\penalty0
  (5):\penalty0 773--815, September 2000.

\bibitem[Ragan-Kelley et~al.(2012)Ragan-Kelley, Adams, Paris, Levoy,
  Amarasinghe, and Durand]{halide_12}
Ragan-Kelley, J., Adams, A., Paris, S., Levoy, M., Amarasinghe, S., and Durand,
  F.
\newblock Decoupling algorithms from schedules for easy optimization of image
  processing pipelines.
\newblock \emph{ACM Trans. Graph.}, 31\penalty0 (4):\penalty0 32:1--32:12, July
  2012.
\newblock ISSN 0730-0301.

\bibitem[Rahman et~al.(2010)Rahman, Pouchet, and Sadayappan]{rahman2010neural}
Rahman, M., Pouchet, L.-N., and Sadayappan, P.
\newblock Neural network assisted tile size selection.
\newblock In \emph{International Workshop on Automatic Performance Tuning
  (IWAPT’2010). Berkeley, CA: Springer Verlag}, 2010.

\bibitem[Smith \& Topin(2017)Smith and
  Topin]{DBLP:journals/corr/abs-1708-07120}
Smith, L.~N. and Topin, N.
\newblock Super-convergence: Very fast training of residual networks using
  large learning rates.
\newblock \emph{CoRR}, abs/1708.07120, 2017.
\newblock URL \url{http://arxiv.org/abs/1708.07120}.

\bibitem[{Trifunovic} et~al.(2009){Trifunovic}, {Nuzman}, {Cohen}, {Zaks}, and
  {Rosen}]{5260526}
{Trifunovic}, K., {Nuzman}, D., {Cohen}, A., {Zaks}, A., and {Rosen}, I.
\newblock Polyhedral-model guided loop-nest auto-vectorization.
\newblock In \emph{2009 18th International Conference on Parallel Architectures
  and Compilation Techniques}, pp.\  327--337, 2009.

\bibitem[Trifunovic et~al.(2010)Trifunovic, Cohen, Edelsohn, Li, Grosser,
  Jagasia, Ladelsky, Pop, Sjodin, and Upadrasta]{trifunovic_graphite_2010}
Trifunovic, K., Cohen, A., Edelsohn, D., Li, F., Grosser, T., Jagasia, H.,
  Ladelsky, R., Pop, S., Sjodin, J., and Upadrasta, R.
\newblock {GRAPHITE} two years after: First lessons learned from {Real-World}
  polyhedral compilation, January 2010.

\bibitem[Vasilache et~al.(2018)Vasilache, Zinenko, Theodoridis, Goyal, DeVito,
  Moses, Verdoolaege, Adams, and Cohen]{Vasilache2018TensorCF}
Vasilache, N., Zinenko, O., Theodoridis, T., Goyal, P., DeVito, Z., Moses,
  W.~S., Verdoolaege, S., Adams, A., and Cohen, A.
\newblock Tensor comprehensions: Framework-agnostic high-performance machine
  learning abstractions.
\newblock \emph{CoRR}, abs/1802.04730, 2018.

\bibitem[Wolf \& Lam(1991)Wolf and Lam]{wolf1991loop}
Wolf, M.~E. and Lam, M.~S.
\newblock A loop transformation theory and an algorithm to maximize
  parallelism.
\newblock \emph{IEEE transactions on parallel and distributed systems},
  2\penalty0 (4):\penalty0 452--471, 1991.

\bibitem[Yang et~al.(2019)Yang, Dai, Yang, Carbonell, Salakhutdinov, and
  Le]{DBLP:journals/corr/abs-1906-08237}
Yang, Z., Dai, Z., Yang, Y., Carbonell, J.~G., Salakhutdinov, R., and Le, Q.~V.
\newblock Xlnet: Generalized autoregressive pretraining for language
  understanding.
\newblock \emph{CoRR}, abs/1906.08237, 2019.
\newblock URL \url{http://arxiv.org/abs/1906.08237}.

\end{thebibliography}
\bibliographystyle{SysML21/mlsys2021}


\appendix
 \clearpage

\section{Appendix}

\vspace{-0.25cm}
\subsection{Model Architecture Details and Training Methodology}\label{modelDetails}
\vspace{-0.25cm}
The computation embedding layer is a fully connected multilayer perceptron (MLP), feedforward neural network that takes 1235-dimensional computation vectors and generates 180-dimensional embeddings. We use 3 intermediate layers of 600, 350, 200 neurons respectively. The output of each layer is transformed by the \emph{ELU} function and fed to a dropout layer with a dropout probability of 0.225, and then passed to the next layer. This succession of the activation function and the dropout layer is applied to all the neural networks of this model. The two LSTM cells in the loop embedding unit have identical input and hidden vector sizes that correspond to the output of the computation embedding layer (180). The feedforward neural network that maps the concatenated hidden vectors to 180-dimensional loop embedding have one intermediate layer of size 200. The regression layer that maps the \emph{program embedding} vector to a speedup value, has two intermediate layers with 200 and 180 neurons. 

We implemented our model in PyTorch \cite{NEURIPS2019_9015} (0.4.1.post2). All model parameters are learnable from the computation embedding layer to the regression layer by way of the recursive loop embedding layer. For the loss function, we used MAPE, a normalized metric based on $L_{1}$. This loss function is suitable for speedup prediction because the target value is positive by design. In addition, the function motivates the model to be equitably accurate in the wide range of speedups we have. The Glorot initialization \cite{glorotunderstanding} is adopted for all weights of the model. We train our model using AdamW \cite{DBLP:journals/corr/abs-1711-05101} with a weight decay coefficient of 0.0075. The learning rate is scheduled by the One Cycle Policy \cite{DBLP:journals/corr/abs-1708-07120} with a maximum learning rate of 0.001. The other optimizer parameters are left on their default values. The best accuracy is achieved after about 700 epochs of training. The training set is processed in batches of 32 data points. Each batch is formed by code transformations belonging to the same algorithm. The rationale for this grouping is that it is faster to operate on data points having the same tree structure.


\vspace{-0.25cm}
\subsection{Benchmark Sizes and Parameters} 
\vspace{-0.25cm}

Table~\ref{tab:benchmark-sizes} summarizes the benchmarks' input sizes and parameters used in Section~\ref{sec:evaluation}.

\begin{table}[h]
\centering
\begin{tabular}{@{}ll@{}}
\toprule
Benchmark & Input size and parameters \\ \midrule
box blur & $3\times1024\times1024$ \\[2mm]
conv + relu & \begin{tabular}[c]{@{}l@{}}batch size: 8\\ input size: $1024\times1024\times3$\\ kernel size: $3\times3$\\ output features: 2\end{tabular} \\[8mm]
convolution & \begin{tabular}[c]{@{}l@{}}batch size: 8\\ input size: $1024\times1024\times3$\\ kernel size: $3\times3$\\ output features: 2\end{tabular} \\[8mm]
cvtcolor & $3\times1024\times1024$ \\
doitgen & $256\times256\times128 ,\  256\times256$ \\
heat2d & $1024\times1024$ \\
heat3d & $770\times898\times1024$ \\
jacobi2d & $130\times1024$ \\
mvt & $1024\times1024$ \\
seidel2d & $256\times256$ \\ \bottomrule
\end{tabular}
\caption{Benchmarks input sizes and parameters}
\label{tab:benchmark-sizes}
\end{table}

\vspace{-0.25cm}
\subsection{More Detailed Evaluation} 
\vspace{-0.25cm}

\begin{figure}[h!]
  \centering
    \includegraphics[width=\columnwidth]{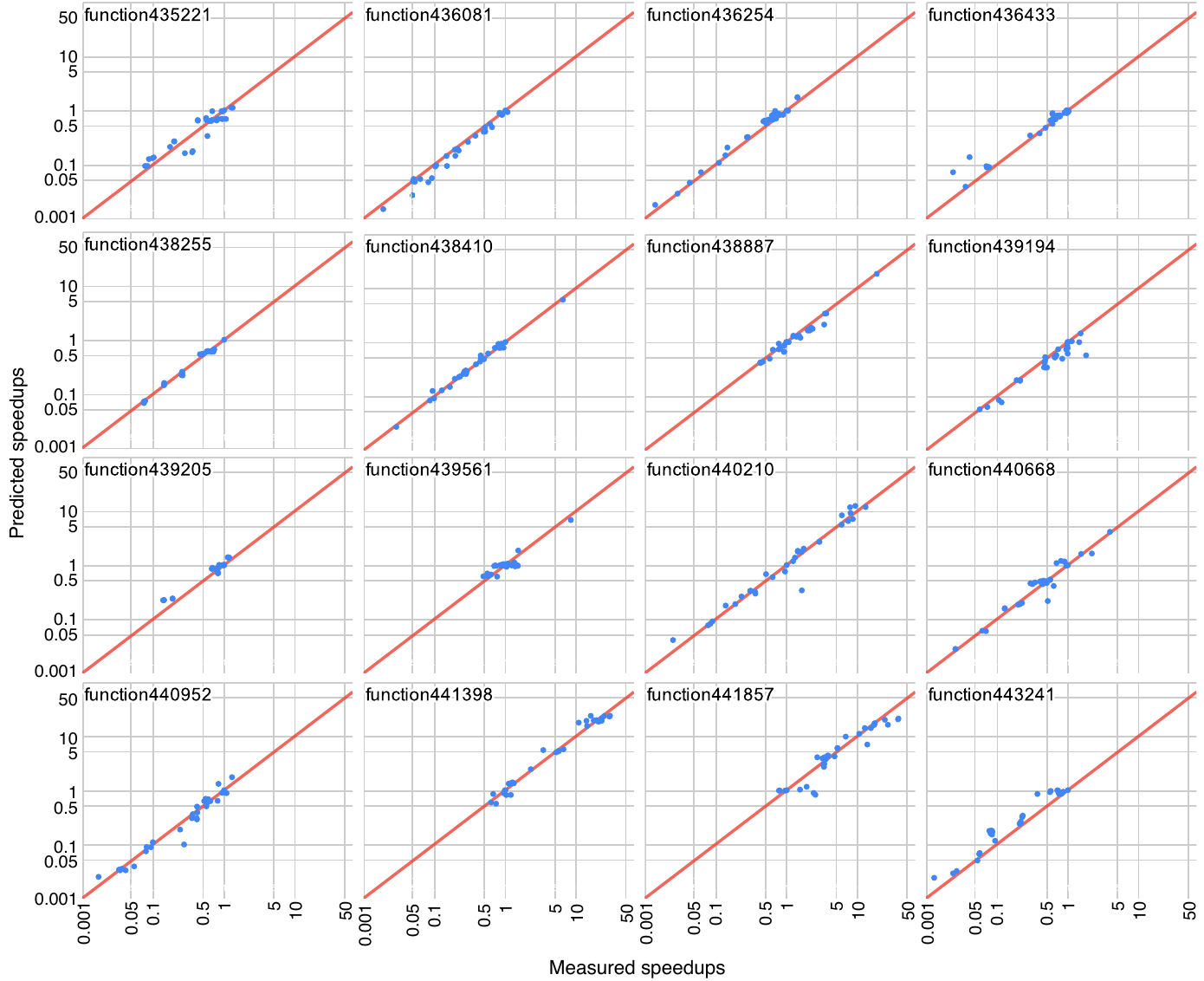}
  \caption{Measured vs predicted speedup on $16$ random programs from the test set, each blue dot represents a code transformation with respect to its measured speedup and its predicted speedup.}
  \label{fig:gridRealVPred}
  \vspace{-0.25cm}
\end{figure}
Figure~\ref{fig:gridRealVPred} gives an overview of the correlation between the predicted and measured speedups over $16$ programs randomly selected from the test set. Each chart represents the $32$ random transformations applied on each program with blue dots, the closer a blue dot is to the red line the lower the prediction error is. This figure shows that the cost model's predictions fit well the distribution and the range of the speedups and does not just predict an average value for each program.

\end{document}